\documentstyle[twoside,amssymb]{article}
\newcommand{\be}{\begin{equation}}
\newcommand{\ee}{\end{equation}}
\newcommand{\bear}{\begin{eqnarray}}
\newcommand{\eear}{\end{eqnarray}}
\catcode`\@=11
\long\def\@makefntext#1{
\protect\noindent \hbox to 3.2pt {\hskip-.9pt  
$^{{\eightrm\@thefnmark}}$\hfil}#1\hfill}		

\def\@makefnmark{\hbox to 0pt{$^{\@thefnmark}$\hss}}	
	
\def\ps@myheadings{\let\@mkboth\@gobbletwo
\def\@oddhead{\hbox{}
\rightmark\hfil\eightrm\thepage}   
\def\@oddfoot{}\def\@evenhead{\eightrm\thepage\hfil
\leftmark\hbox{}}\def\@evenfoot{}
\def\sectionmark##1{}\def\subsectionmark##1{}}

\oddsidemargin=\evensidemargin
\newcounter{sectionc}\newcounter{subsectionc}\newcounter{subsubsectionc}
\renewcommand{\section}[1] {\vspace{12pt}\addtocounter{sectionc}{1} 
\setcounter{subsectionc}{0}\setcounter{subsubsectionc}{0}\noindent 
	{\tenbf\thesectionc. #1}\par\vspace{5pt}}
\renewcommand{\subsection}[1] {\vspace{12pt}\addtocounter{subsectionc}{1} 
	\setcounter{subsubsectionc}{0}\noindent 
	{\bf\thesectionc.\thesubsectionc. {\kern1pt \bfit #1}}\par\vspace{5pt}}
\renewcommand{\subsubsection}[1] {\vspace{12pt}\addtocounter{subsubsectionc}{1}
	\noindent{\tenrm\thesectionc.\thesubsectionc.\thesubsubsectionc.
	{\kern1pt \tenit #1}}\par\vspace{5pt}}
\newcommand{\nonumsection}[1] {\vspace{12pt}\noindent{\tenbf #1}
	\par\vspace{5pt}}

\newcounter{appendixc}
\newcounter{subappendixc}[appendixc]
\newcounter{subsubappendixc}[subappendixc]
\renewcommand{\thesubappendixc}{\Alph{appendixc}.\arabic{subappendixc}}
\renewcommand{\thesubsubappendixc}
	{\Alph{appendixc}.\arabic{subappendixc}.\arabic{subsubappendixc}}

\renewcommand{\appendix}[1] {\vspace{12pt}
        \refstepcounter{appendixc}
        \setcounter{figure}{0}
        \setcounter{table}{0}
        \setcounter{lemma}{0}
        \setcounter{theorem}{0}
        \setcounter{corollary}{0}
        \setcounter{definition}{0}
        \setcounter{equation}{0}
        \renewcommand{\thefigure}{\Alph{appendixc}.\arabic{figure}}
        \renewcommand{\thetable}{\Alph{appendixc}.\arabic{table}}
        \renewcommand{\theappendixc}{\Alph{appendixc}}
        \renewcommand{\thelemma}{\Alph{appendixc}.\arabic{lemma}}
        \renewcommand{\thetheorem}{\Alph{appendixc}.\arabic{theorem}}
        \renewcommand{\thedefinition}{\Alph{appendixc}.\arabic{definition}}
        \renewcommand{\thecorollary}{\Alph{appendixc}.\arabic{corollary}}
        \renewcommand{\theequation}{\Alph{appendixc}.\arabic{equation}}
        \noindent{\tenbf Appendix \theappendixc #1}\par\vspace{5pt}}
\newcommand{\subappendix}[1] {\vspace{12pt}
        \refstepcounter{subappendixc}
        \noindent{\bf Appendix \thesubappendixc. {\kern1pt \bfit #1}}
	\par\vspace{5pt}}
\newcommand{\subsubappendix}[1] {\vspace{12pt}
        \refstepcounter{subsubappendixc}
        \noindent{\rm Appendix \thesubsubappendixc. {\kern1pt \tenit #1}}
	\par\vspace{5pt}}

\newcommand{\textlineskip}{\baselineskip=13pt}
\newcommand{\smalllineskip}{\baselineskip=10pt}

\def\eightcirc{
\begin{picture}(0,0)
\put(4.4,1.8){\circle{6.5}}
\end{picture}}
\def\eightcopyright{\eightcirc\kern2.7pt\hbox{\eightrm c}} 

\newcommand{\copyrightheading}[1]
	{\vspace*{-2.5cm}\smalllineskip{\flushleft
	{\footnotesize International Journal of Modern Physics A, #1}\\
	{\footnotesize $\eightcopyright$\, World Scientific Publishing
	 Company}\\
	 }}

\def\abstracts#1#2#3{{
	\centering{\begin{minipage}{4.5in}\baselineskip=10pt\footnotesize
	\parindent=0pt #1\par 
	\parindent=15pt #2\par
	\parindent=15pt #3
	\end{minipage}}\par}} 

\renewenvironment{thebibliography}[1]
	{\frenchspacing
	 \ninerm\baselineskip=11pt
	 \begin{list}{\arabic{enumi}.}
	{\usecounter{enumi}\setlength{\parsep}{0pt}
	 \setlength{\leftmargin 12.7pt}{\rightmargin 0pt} 
	 \setlength{\itemsep}{0pt} \settowidth
	{\labelwidth}{#1.}\sloppy}}{\end{list}}

\newcounter{itemlistc}
\newcounter{romanlistc}
\newcounter{alphlistc}
\newcounter{arabiclistc}

\newcommand{\fcaption}[1]{
        \refstepcounter{figure}
        \setbox\@tempboxa = \hbox{\footnotesize Fig.~\thefigure. #1}
        \ifdim \wd\@tempboxa > 5in
           {\begin{center}
        \parbox{5in}{\footnotesize\smalllineskip Fig.~\thefigure. #1}
            \end{center}}
        \else
             {\begin{center}
             {\footnotesize Fig.~\thefigure. #1}
              \end{center}}
        \fi}

\newcommand{\tcaption}[1]{
        \refstepcounter{table}
        \setbox\@tempboxa = \hbox{\footnotesize Table~\thetable. #1}
        \ifdim \wd\@tempboxa > 5in
           {\begin{center}
        \parbox{5in}{\footnotesize\smalllineskip Table~\thetable. #1}
            \end{center}}
        \else
             {\begin{center}
             {\footnotesize Table~\thetable. #1}
              \end{center}}
        \fi}

\def\@citex[#1]#2{\if@filesw\immediate\write\@auxout
	{\string\citation{#2}}\fi
\def\@citea{}\@cite{\@for\@citeb:=#2\do
	{\@citea\def\@citea{,}\@ifundefined
	{b@\@citeb}{{\bf ?}\@warning
	{Citation `\@citeb' on page \thepage \space undefined}}
	{\csname b@\@citeb\endcsname}}}{#1}}

\newif\if@cghi
\def\cite{\@cghitrue\@ifnextchar [{\@tempswatrue
	\@citex}{\@tempswafalse\@citex[]}}
\def\citelow{\@cghifalse\@ifnextchar [{\@tempswatrue
	\@citex}{\@tempswafalse\@citex[]}}
\def\@cite#1#2{{$\null^{#1}$\if@tempswa\typeout
	{IJCGA warning: optional citation argument 
	ignored: `#2'} \fi}}

\def\pmb#1{\setbox0=\hbox{#1}
	\kern-.025em\copy0\kern-\wd0
	\kern.05em\copy0\kern-\wd0
	\kern-.025em\raise.0433em\box0}

\def\fnt#1#2{\footnotetext{\kern-.3em
	{$^{\mbox{\scriptsize #1}}$}{#2}}}

\def\fpage#1{\begingroup
\voffset=.3in
\thispagestyle{empty}\begin{table}[b]\centerline{\footnotesize #1}
	\end{table}\endgroup}

\def\runninghead#1#2{\pagestyle{myheadings}
\markboth{{\protect\footnotesize\it{\quad #1}}\hfill}
{\hfill{\protect\footnotesize\it{#2\quad}}}}
\font\tenrm=cmr10
\font\tenit=cmti10 
\font\tenbf=cmbx10
\font\bfit=cmbxti10 at 10pt
\font\ninerm=cmr9

\font\eightrm=cmr8

\def\qed{\hbox{${\vcenter{\vbox{			
   \hrule height 0.4pt\hbox{\vrule width 0.4pt height 6pt
   \kern5pt\vrule width 0.4pt}\hrule height 0.4pt}}}$}}

\begin{document}

\runninghead{Resolution of the  Strong CP Problem} {}

\normalsize\textlineskip
\thispagestyle{empty}
\setcounter{page}{1}

\copyrightheading{}			

\vspace*{0.88truein}

\fpage{1}
\centerline{\bf Resolution of the  Strong CP Problem}
\vspace*{0.035truein}
\centerline{\bf }
\vspace*{0.37truein}
\centerline{\footnotesize TAEKOON LEE\footnote{tlee@muon.kaist.ac.kr}}
\vspace*{0.015truein}
\centerline{\footnotesize\it Physics Department, KAIST, Daejon 305-701, Korea}
\baselineskip=10pt

\vspace*{0.21truein}
\abstracts{It is shown that the quark mass aligns  QCD $\theta$ vacuum in
such a way that the strong CP is conserved, 
resolving the strong CP problem.}{}{}

\vspace*{12pt}			

\vspace*{1pt}\textlineskip	
\vspace*{-0.5pt}
\noindent
Quantum chromodynamics (QCD) is well established as the fundamental theory 
for the strong interactions. However, there has been a persistent puzzle
with QCD, namely the smallness of the strong CP violation. The most general
QCD Lagrangian is given in the form:
\begin{equation}
{\cal L} = {\cal L}_0 +{\cal L}_{\theta_0} +{\cal L}_{m},
\label{e1}
\ee
where
\bear
{\cal L}_0&=& -\frac{1}{4} F_{\mu\nu}^a F^{a\mu\nu} \nonumber \\
          &&+i \sum_i\left[
 \bar{\psi}_{L,i}(\not\!\partial+i g \not\!\! A^a T^a)
 \psi_{L,i} + (L \rightarrow R)\right],
 \nonumber \\
 {\cal L}_{\theta_0}&=& \frac{\theta_0}{32\pi^2} 
 F_{\mu\nu}^a\tilde{F}^{a\mu\nu},
 \nonumber \\
 {\cal L}_m&=& \sum_{ij} m_{ij} \bar{\psi}_{L,i} \psi_{R,j} +{\mathrm H.c.}
\label{e2}
\eear
As usual $F_{\mu\nu}^a$ denotes the field strength tensor for the gluons
$A_\mu^a$, $g$  and $\theta_0$ are coupling constants,
and $\psi_{L(R),i} =\frac{1}{2} 
(1\mp\gamma_5)\psi_i, i=1,\cdots, {\mathrm N}_f$,
denote the ${\mathrm N}_f$  quark flavors.
The quark mass $m_{ij}$ can be an arbitrary complex matrix, but using the
${\mathrm SU}({\mathrm N}_f)_L\times {\mathrm SU}({\mathrm N}_f)_R$ 
chiral symmetry of
${\cal L}_0$ can be written without loss of generality as 
\be
m_{ij}= m_i \delta_{ij} e^{i\delta},
\label{e3}
\ee
where $m_i$ are real and positive,  and $\delta$ is a constant phase.
As well known, the axial ${\mathrm U(1)}$ anomaly allows one to shift the
phase $\delta$ into ${\cal L}_{\theta_0}$ and vice versa \cite{fugi}. 
This property
can be used to remove ${\cal L}_{\theta_0}$, and write the QCD Lagrangian 
(\ref{e1}) as
\be
\tilde{{\cal L}} = {\cal L}_0 + {\cal L'}_m,
\label{e4}
\ee
where 
\be
{\cal L'}_m= \sum_i m_i\left( e^{i\bar\delta} \bar{\psi}_{L,i} \psi_{R,i}
+{\mathrm H.c.}\right)
\label{e-5}
\ee
with
$\bar\delta=\delta +\theta_0/{\mathrm N}_f$.
The two theories (\ref{e1}) and (\ref{e4}) are completely equivalent.

Apparently, with nonzero $\bar\delta$ the theory (\ref{e4}) appears to
break CP.
The experimental bound on neutron electric dipole moment suggests $\bar\delta$
 be extremely small: $\bar\delta < 10^{-9}$ \cite{baluni,cvvw}.
Why is $\bar\delta$ so small? This is the strong CP problem (for review
see \cite{kc}).

A small number, unless  protected by  symmetry, demands for its smallness
a {\it natural} explanation that does not require fine tuning.
Among the various solutions proposed for the problem, the most popular is the
Peccei-Quinn mechanism \cite{pq}. 
It predicts a very light and weakly interacting
particle, the axion \cite{ww,kd}.
Axion, however, has not been observed, and the
current window for its mass is quite narrow, ranging from $10^{-6}{\mathrm eV}$
to $10^{-3}{\mathrm eV}$ \cite{turner}.

In this Letter we show that there is a mechanism within QCD that renders
the strong CP conservation {\it automatic}.
The mechanism we propose relies  on the QCD $\theta$ vacua and the
isospin singlet, pseudoscalar meson $\eta'$.

Let us assume for the moment that ${\cal L'}_m=0$, i.e., quarks are massless
(massless QCD).
Then the axial U(1) is broken only by the Adler-Bell-Jackiw anomaly 
\cite{anomaly}
\be
\partial_\mu J_5^\mu = \frac{g^2{\mathrm N}_f}
{16\pi^2}  F_{\mu\nu}^a\tilde{F}^{a\mu\nu}=
-\partial_\mu K^\mu,
\ee
where
\be
J_5^\mu= \sum_i \bar{\psi}_i\gamma^\mu\gamma_5\psi_i
\ee
is the axial current, and 
\be
K^\mu= -\frac{g^2{\mathrm N}_f}
{16\pi^2} \epsilon^{\mu\alpha\beta\gamma}A^a_\alpha
(F^a_{\beta\gamma}-\frac{1}{3} \epsilon^{abc}A^b_\beta A^c_\gamma)
\ee
is the gauge-dependent topological current.
This anomaly equation gives rise to a gauge-dependent conserved
charge $\tilde{Q}_5$:
\be
\tilde{Q}_5=\int (J_5^0 +K^0)(\vec{x},t)d^3\vec{x}.
\label{e9}
\ee

The symmetry of the massless QCD under the rotation
\be
U_5(\theta)=e^{i\theta\tilde{Q}_5}
\label{e10}
\ee
can be used to construct a continuum of equivalent vacua.
Let $|\Omega_0\rangle$ be a vacuum of the massless QCD.
Since at low energies the chiral symmetry
${\mathrm SU}({\mathrm N}_f)_L\times {\mathrm SU}({\mathrm N}_f)_R$ is
spontaneously broken to ${\mathrm SU}({\mathrm N}_f)_{L+R}$,
$|\Omega_0\rangle$
must belong to the chiral vacua
$[{\mathrm SU}({\mathrm N}_f)_L\times
{\mathrm SU}({\mathrm N}_f)_R]/{\mathrm SU}({\mathrm N}_f)_{L+R}$.
Now, the state $|\theta\rangle$ defined by
\be
|\theta\rangle=U_5(\theta)|\Omega_0\rangle
\label{e11}
\ee
can also be a vacuum, equivalent to $|\Omega_0\rangle$,
since $\tilde{Q}_5$ commutes
with the Hamiltonian of the massless QCD. Note that $|\theta\rangle$
does not belong to the physical Hilbert space built on the vacuum
$|\Omega_0\rangle$
because $\tilde{Q}_5$ is not gauge invariant. Therefore,
there is a continuum of
vacua, the $\theta$ vacua \cite{ls,cdg}.
Hence, the continuum of the vacua  for the  massless QCD is 
 given as the product of
the $\theta$ vacua and the chiral vacua.
Any point in the continuum can be taken as
the vacuum, and the physics is oblivious of the particular choice.

Let us now consider the transformation of $\eta'$ 
under the $U_5(\theta)$.
Since $\eta'$ couples to the axial current $J_5^\mu$, $\exp(i\eta'/f_{\eta'})$
transforms as $\sum_i \bar{\psi}_{L,i}\psi_{R,i}$ under $U_5(\theta)$.
This gives
\be
U_5(\theta)^\dagger \eta' U_5(\theta)=\eta' + 2 f_{\eta'} \theta
\label{e12}\ee
because
\be
U_5(\theta)^\dagger\sum_i \bar{\psi}_{L,i}\psi_{R,i}
U_5(\theta)= e^{2 i \theta}
\sum_i \bar{\psi}_{L,i}\psi_{R,i}.
\label{transLR}
\ee
The $\eta'$ decay constant $f_{\eta'}$ is a constant of dimension one.
Using Eq. \ (\ref{e12})
we obtain
\bear
\langle \theta+\delta\theta|\eta'|\theta+\delta\theta\rangle &=&
\langle \theta| U_5(\delta\theta)^\dagger \eta' U_5(\delta\theta)
|\theta\rangle
\nonumber \\
&=& \langle \theta|\eta'|\theta\rangle +2 f_{\eta'} \delta\theta.
\label{e13}
\eear
And using Eq. \ (\ref{transLR}) we can also write the quark condensates in
the vacuum $|\theta\rangle$ \footnote{This vacuum 
should really be regarded as a point in the continuum of
the vacua for the massless QCD; for notational convenience, only the
$\theta$ component is made explicit whereas  the chiral component is
suppressed.} as
\be
\langle\theta|\bar{\psi}_{L,i}\psi_{R,j}|\theta\rangle=|\Delta|
\Sigma^{(0)}_{ij} e^{i\Phi_0},
\label{condensates}
\ee
where $\Phi_0=2\theta +\phi_0$, with $\phi_0$ being
a constant phase, and $\Delta$ is a constant of dimension
three while $\Sigma^{(0)} \in {\mathrm SU}({\mathrm N}_f)$.
Thus the phase of the quark condensates tells which $\theta$ vacuum the system
is in.

We now have all the tools to present our mechanism.
For our purpose we can ignore heavy quarks
and keep only the first three flavors (${\mathrm N}_f=3$).
Let us now turn on a
{\it small} quark mass, so that ${\cal L'}_m$ can be regarded as a
perturbation to ${\cal L}_0$. Let $|\Omega_m\rangle$ be the vacuum of
the theory (\ref{e4}). 
Since the isospin breaking by the quark mass is small we can write the
quark condensates in leading order as
\be
\langle\Omega_m|\bar{\psi}_{L,i}\psi_{R,j}|\Omega_m\rangle=|\Delta|
\Sigma^{(m)}_{ij} e^{i\Phi_m},
\label{condensates2}
\ee
where $\Sigma^{(m)} \in {\mathrm SU}(3)$ and $\Phi_m$ is a constant phase.

Our crucial observation is that the phase $\Phi_m$ as well as $\Sigma^{(m)}$
must be determined dynamically, and consequently that they will depend on
the quark mass ($m_i$ as well as the phase $\bar\delta$). It is not
surprising to expect this,
since the vacuum $|\Omega_m\rangle$ would in general depend on the
parameters of the theory. It is, in fact, well known that $\Sigma^{(m)}$ is
determined dynamically through
the Dashen's theorem \cite{dashen},
and is dependent on the quark mass. This mechanism is called chiral vacuum
alignment. We shall show that $\Phi_m$ is also determined dynamically.

Note that because the phase of the  quark condensates before the perturbation
${\cal L}'_m$ is turned on  is dependent on the
$\theta$ vacuum chosen  (Eq. \ (\ref{condensates})),
the dynamical determination
of the phase $\Phi_m$ implies dynamical selection of the  $\theta$ vacuum
by the quark mass---$\theta$ vacuum alignment---exactly in the 
manner the quark mass selects the chiral vacuum.

Before we show that the quark mass actually aligns the $\theta$ vacuum, we
observe that with no $\theta$ vacuum alignment the theory (\ref{e4})
is inherently ambiguous, in that the physics cannot be determined completely
in terms of the parameters of the theory.
To see this, we shall assume that there is no $\theta$ vacuum
alignment, and that  $|\theta\rangle$ with the quark condensates 
(\ref{condensates}) was the vacuum of the massless
QCD before the quark mass was turned on.
To correctly describe the quark mass effects one must first
find the true chiral vacuum of the theory \cite{dashen,baluni}.
Dashen's theorem requires the potential, which lifts the degeneracy
of the chiral vacua,
\bear
V(\Sigma^{(0)})&=&\langle\theta|\delta {\cal H}|\theta\rangle \nonumber \\
&=&-\langle\theta| {\cal L}'_m|\theta\rangle \nonumber \\
&=&- 2|\Delta| {\mathrm Re}\left[ e^{i(\bar\delta+\Phi_0)}\sum_i m_i
\Sigma^{(0)}_{ii}
\right],
\eear
where $\delta {\cal H}$ is the perturbation to the Hamiltonian density,
be minimized over the variable $\Sigma^{(0)}$.
In general the true chiral vacuum that minimizes $V(\Sigma^{(0)})$,
will depend on the phase $\bar\delta+\Phi_0$, and consequently
the physics, in particular CP violation, that depends on $\bar\delta$
will arise only through the combined phase $\bar\delta+\Phi_0$.
Since $\Phi_0$ is not a parameter that appears in the Lagrangian
(\ref{e4}), and depends entirely on our choice of the $\theta$ vacuum,
the physics is  ambiguous.\footnote{It is worthwhile to note that
the generally accepted, CP violating effective Lagrangian by Baluni
\cite{baluni}
is in fact ambiguous. In his derivation of the effective Lagrangian
he erroneously put $\Phi_0=0,\pi$ and $\Sigma^{(0)}_{ij}=\delta_{ij}$
by requiring that the quark condensates (\ref{condensates})
be real so that the vacuum is
CP even. However, we note that there is no reason to require the vacuum be
CP symmetric, and furthermore the apparent CP violation by complex
quark condensates in massless QCD
is a fictitious one because the physics is independent on the choice of
a particular $\theta$ vacuum. Therefore, the Baluni's effective 
Lagrangian must
depend on the arbitrary phase $\Phi_0$. Note that even if one
accepts his argument there is still an ambiguity of choosing
 $\Phi_0$ between the two values, 0 and $\pi$.}

This ambiguity  can also be demonstrated in the two dimensional
Schwinger model \cite{schwinger}:
\bear
{\cal L}&=& -\frac{1}{4} F_{\mu\nu} F^{\mu\nu} +i \left[
 \bar{\psi}_{L}(\not\!\partial+i e \not\!\! A)
 \psi_{L} + (L \rightarrow R)\right] \nonumber \\
 &&+m( e^{i\delta }\bar{\psi}_{L} \psi_{R} +{\mathrm H.c.}),
\label{schwinger}
\eear
where $A_\mu$ is the U(1) gauge field,  $m$ is the fermion mass,
and $\delta$ is a constant phase. In this model the axial U(1) is
anomalous, and therefore a continuum of $\theta$ vacua can be
constructed in a similar manner as in the massless QCD \cite{ls}.
Like ${\cal L}_{\theta_0}$ in QCD, a topological term
$\epsilon_{\mu\nu}F^{\mu\nu}$, where $\epsilon_{\mu\nu}$ is a constant
antisymmetric tensor, may be added
to the Lagrangian (\ref{schwinger}), but as before 
it can be removed by a proper
axial U(1) rotation of the fermion field. The
phase $\delta$ then plays the role of $\bar\delta$ in ${\cal L}'_m$ of QCD.

The easiest way to see the ambiguity is by bosonization of 
the Lagrangian (\ref{schwinger}) using the standard rule \cite{coleman}
\bear
&&\bar\psi i\not\!\partial \psi= \frac{1}{2} (\partial_\mu \sigma)^2
\nonumber \\
&&\bar\psi \gamma^\mu \psi=\frac{1}{\sqrt{\pi}}
\epsilon^{\mu\nu}\partial_\nu\sigma
\nonumber \\
&&\bar\psi_L\psi_R= \frac{e}{2}\exp(i \sqrt{4\pi}\sigma).
\label{rule}
\eear
Note that this  bosonization has an inherent ambiguity:
the freedom to shift $\sigma \to\sigma+ \sigma_0$, where $\sigma_0$ is an
arbitrary constant.
Because this ambiguity affects only the last equation in (\ref{rule}),
and consequently the fermion condensate,
it is easy
to see that the ambiguity corresponds to a particular selection of
the $\theta$ vacuum.
With this freedom taken into account, the bosonized Lagrangian is given as:
\bear
{\cal L}&=& -\frac{1}{4} F_{\mu\nu} F^{\mu\nu}
+\frac{1}{2}(\partial_\mu\sigma)^2 -
\frac{e}{\sqrt{\pi}}\sigma\epsilon^{\mu\nu}\partial_\mu A_\nu\nonumber\\
&&+ m \cos[\delta +\sqrt{4\pi}(\sigma+\sigma_0)],
\label{bosonized}
\eear
which clearly shows that the $\delta$ dependence occurs through
$\delta+\sqrt{4\pi}\sigma_0$, and so ambiguous because it depends
on an arbitrary parameter.
Now, what is the implication of this ambiguity? It implies that
$\sigma_0$, and accordingly the $\theta$ vacuum,
must be determined dynamically because the theory
(\ref{schwinger}) cannot be ambiguous.  It will be shown shortly that
the $\theta$ vacuum alignment  determines
$\sigma_0$ dynamically and removes this ambiguity.

We now show that the quark mass in fact induces $\theta$ vacuum alignment.
As before 
let us assume that $|\theta\rangle$ was the vacuum of the massless QCD
before the quark mass was turned on, and that the quark condensates were given
by (\ref{condensates}).

The $\eta'$ in the vacuum $|\theta\rangle$ can be
described by an effective Lagrangian:
\be
{\cal L}_{\eta'} =\frac{1}{2} (\partial_\mu\eta')^2-\frac{1}{2}
m_{\eta'}^2\eta'^2 +
{\mathrm interactions},
\label{effective}
\ee
where the ignored terms involve cubic or higher powers of the fields, and
$m_{\eta'}$ is the $\eta'$ mass in massless QCD induced by the nonvanishing
topological susceptibility \cite{wtn}.
With this Lagrangian, the $\eta'$ satisfies
\be
\langle\theta|\eta'|\theta\rangle =0.
\label{vev}
\ee
Upon turning on the quark mass, 
${\cal L'}_m$ induces the following interaction to ${\cal L}_{\eta'}$:
\bear
&&\sum_i m_i\,\,|\Delta|\Sigma^{(0)}_{ii} e^{i(\bar\delta+\Phi_0)} 
e^{i\eta'/f_{\eta'}} +{\mathrm H.c.}
\nonumber \\
&&\,\,\,\,= {\mathrm const.} - 2 |M \Delta| \left[\sin(\bar\delta+
\Phi_0+\phi_{\Sigma^{(0)}})
(\eta'/f_{\eta'})\right. \nonumber \\
&&\,\,\,\,+ \frac{1}{2}
\cos(\bar\delta+\Phi_0+\phi_{\Sigma^{(0)}})(\eta'/f_{\eta'})^2]
+O(\eta'^3),
\label{mass-eta-int}
\eear
where $|M|$ and $\phi_{\Sigma^{(0)}}$ are defined through
\be
\sum_i m_i \Sigma^{(0)}_{ii}=|M| e^{i\phi_{\Sigma^{(0)}}}.
\ee
The effective Lagrangian (\ref{effective}) 
is thus modified by the quark mass as
\bear
\tilde{\cal L}_{\eta'}& =&\frac{1}{2} (\partial_\mu\eta')^2-\frac{1}{2}
\tilde{m}_{\eta'}^2\eta'^2 \nonumber \\ 
&&-  2 |M \Delta|\sin(\bar\delta+\Phi_0+\phi_{\Sigma^{(0)}})(\eta'/f_{\eta'})
\nonumber \\
&&+{\mathrm interactions},
\label{ee24}
\eear
where
\be
\tilde{m}_{\eta'}^2=m_{\eta'}^2 +2|M\Delta|
\cos(\bar\delta+\Phi_0+\phi_{\Sigma^{(0)}})/f_{\eta'}^2.
\label{eta-mass}\ee
With the presence of a term linear in $\eta'$ (for $\theta$ satisfying
$\sin(\bar\delta+\Phi_0+\phi_{\Sigma^{(0)}}) \neq 0$)
this equation
shows that
the quark mass exerts a force on $\eta'$ and
pushes it away from its stable position at $\eta'=0$; and, therefore,
the equation (\ref{vev}) no longer holds.
With the help of  Eq. \ (\ref{e13}) we can then see that
 this shift in the vacuum expectation value of $\eta'$ implies
a realignment of the QCD system  to
a new $\theta$ vacuum. For instance, when the vacuum expectation value
of $\eta'$ shifts from zero to a nonzero value, say $\delta v$, the system
rotates by $\delta\theta=\delta v/2f_{\eta'}$.
The $\theta$ vacuum thus becomes unstable in presence of
the quark mass, and realigns successively to  a new $\theta$ vacuum, 
through the
interaction of $\eta'$ to the quark mass, until the linear term in
(\ref{ee24}) vanishes.
Note that this phenomenon is not much different from
the realignment of a ferromagnetic system
at the introduction of an external magnetic field.

An important aspect of the Lagrangian (\ref{ee24}) is that it should be
regarded as valid
only for an infinitesimal $\eta'$ except when the vacuum $|\theta\rangle$ 
is the stable one. Given a Lagrangian, one would usually 
minimize the potential of the Lagrangian
to find the vacuum and do perturbation around
it to read off particle spectrum and interactions. However, with the
Lagrangian (\ref{ee24}) this would give wrong physics.
For instance, a straightforward minimization of the potential in
(\ref{ee24}) would suggest the $\theta$ vacuum rotate
only by an amount $\delta\theta \propto
|M\Delta|/(\tilde{m}_{\eta'}^2 f_{\eta'}^2) \sin(
\bar\delta+\Phi_0+\phi_{\Sigma^{(0)}})$ whereas according to our
argument above
the system actually should rotate
until $\sin(\bar\delta +\Phi_0+\phi_{\Sigma^{(0)}})=0$ is satisfied; And also
it would give the $\eta'$-mass by the unacceptable
formula (\ref{eta-mass}) which is
ambiguous.
The reason that the usual procedure fails with the Lagrangian (\ref{ee24})
is that the quark mass
induced term (\ref{mass-eta-int}) is linked to the condition (\ref{vev});
As soon as the system aligns to a new $\theta$ vacuum, the Eq. (\ref{vev})
no longer holds, and accordingly the quark mass induced term
 should be modified through the shift in the phase $\Phi_0$.
This makes the effective Lagrangian for $\eta'$ dependent on
 the $\theta$ vacuum.
As an example, when the system rotates from $|\theta\rangle$
to $|\theta+\delta\theta\rangle$, the Lagrangian for
$\eta'$ in the new vacuum $|\theta+\delta\theta\rangle$
is given by (\ref{ee24}), but now with
$\Phi_0=2(\theta+\delta\theta) +\phi_0$. In the usual, spontaneously
broken case, this shift in the phase
can be absorbed by making a shift in the the associated Nambu-Goldstone boson
field, leaving the Lagrangian invariant; consequently,
the vacuum alignment in this case is equivalent to the familiar picture
of the rolling of the Nambu-Goldstone
boson field from an unstable vacuum to the stable one.
However, in the case of $\eta'$, the rotation of the QCD system cannot
be interpreted as the rolling of $\eta'$ because the nonzero $\eta'$ mass
in the massless QCD does not allow one to absorb $\delta\Phi_0$
by making a shift in the $\eta'$ field.
Therefore, physical quantities like the $\eta'$ mass can be 
read off correctly  only from the effective Lagrangian written 
after the system
settled down on the stable $\theta$ vacuum.

One may recall at this moment the general belief that the $\theta$
vacuum is stable
against perturbation. The essential point of the argument for the
$\theta$ vacuum stability 
\cite{cjs} is that there is no Nambu-Goldstone boson associated with
the symmetry  $U_5(\theta)$. In the usual, spontaneously broken
case the symmetry breaking term lifts the degeneracy of the vacua
and also couples to the associated  Nambu-Goldstone bosons. An unstable vacuum
thus decays to the stable one by emitting the Nambu-Goldstone bosons.
Of course, an identical process cannot happen 
in the $\theta$ vacua because there
is no associated Nambu-Goldstone boson. However, we must realize that
the absence of the Nambu-Goldstone boson does not prevent the $\theta$
vacuum from decaying. As we can see in (\ref{mass-eta-int}), the quark mass
term, which breaks the $U_5(\theta)$ symmetry, couples to $\eta'$, 
and thus an unstable
$\theta$ vacuum can decay through the emission of $\eta'$s. In this case,
however, the decay rate of the unstable vacuum would be slower than 
the usual case because of the large $\eta'$ mass.

Now, because the $\theta$ vacuum becomes dynamical in presence
of the quark mass
the true QCD vacuum $|\Omega_m\rangle$ in (\ref{condensates2})
can be  picked up from the continuum of the vacua 
of the massless QCD ($\theta$ vacua times chiral vacua) by minimizing
the potential
\bear
V(\Phi,\Sigma)&=&-\langle\Omega| {\cal L}'_m|\Omega\rangle \nonumber \\
&=& -2|\Delta|
{\mathrm Re}\left[ e^{i(\bar\delta +\Phi)}\sum_i m_i\Sigma_{ii},
\right]
\eear
over the variables $\Phi$ and $\Sigma \in {\mathrm SU(3)}$.
Here $|\Omega\rangle$ denotes a point in the continuum of the vacua of
the massless QCD,
and  the quark condensates in $|\Omega\rangle$ are given by
\be
\langle\Omega|\bar{\psi}_{L,i}\psi_{R,j}|\Omega\rangle=|\Delta|
\Sigma_{ij} e^{i\Phi}.
\ee
Since by definition $m_i$ are positive, it is  trivial to see that the
potential has a  minimum at $\Phi=\Phi_m$, $\Sigma=\Sigma^{(m)}$, where
\be
\Phi_m=-\bar\delta, \,\,\,\Sigma^{(m)}=I.
\label{relation}
\ee
This is the quark-mass dependence of the quark
condensates (\ref{condensates2}) in the true vacuum.
Note that the phase of the quark condensates cancels exactly the
CP violating phase $\bar\delta$ in the QCD Lagrangian. Since the
low energy CP violation occurs only through the combined phase of the
quark condensates and the QCD Lagrangian this shows no CP violation
at low energies.

For low energy QCD Lagrangian, we notice 
\be
\langle\Omega_m|{\cal L}'_m|\Omega_m\rangle= 2 \sum_i m_i |\Delta|
\ee
from Eqs. \ (\ref{condensates2}) and (\ref{relation}),
and hence, as far as low energy physics is concerned,     
the  quark mass term ${\cal L}'_m$
takes the following form in the true vacuum $|\Omega_m\rangle$:
\be
 {\cal L}''_m= \sum_i m_i \bar\psi_i\psi_i
\ee
where the quark fields $\psi_i$ are normalized as
\be
\langle\Omega_m|\bar\psi_{L,i}\psi_{R,j}|\Omega_m\rangle
=|\Delta|\delta_{ij}.
\ee
Therefore, the QCD Lagrangian for low energy physics 
can be written as
\be
{\cal L}_{{\mathrm QCD}} ={\cal L}_0 + {\cal L}''_m,
\ee
which evidently shows that the physics is independent of $\bar\delta$ and CP
is conserved.
This resolves the strong CP problem.

Now, what would be the $\delta$ dependence of the Schwinger model?
For the essentially same reason given for the dynamical alignment
of QCD $\theta$ vacuum, the $\theta$ vacuum alignment by fermion mass term
should occur in the Schwinger model too.
It is then easy to see
the $\theta$ vacuum alignment requires $\sigma_0=-\delta/\sqrt{4\pi}$ in
(\ref{bosonized}),
and so $\delta$ disappears from the theory. The physics is thus
independent of $\delta$.

Finally, we briefly comment on the physical significance
of the CP violating phase
$\bar\delta$ in (\ref{e-5}) at high energies. Although this phase become
irrelevant at low energies, in chirally symmetric phase where
no quark condensation occurs there would be no $\theta$ vacuum alignment,
and so the CP violation by $\bar\delta$  would
become observable.
This then could be a potential source of CP violation that might become 
important in  physics involving CP violation, for example, 
such as electroweak baryogenesis.

\nonumsection{Acknowledgements}
\noindent
This work was supported by BK21 Core Project.

\nonumsection{References}

\end{document}